\documentclass[letterpaper,12pt]{article}
\usepackage{epsfig}
\usepackage{amssymb}
\usepackage{amsfonts}
\usepackage{amsmath}
\usepackage{graphicx}

\newskip\humongous \humongous=0pt plus 1000pt minus 1000pt

\newif\ifdtup

\catcode`\@=11

\@addtoreset{equation}{section}
\def\theequation{\thesection.\arabic{equation}}

\def\@normalsize{\@setsize\normalsize{15pt}\xiipt\@xiipt
\abovedisplayskip 14pt plus3pt minus3pt%
\belowdisplayskip \abovedisplayskip
\abovedisplayshortskip \z@ plus3pt%
\belowdisplayshortskip 7pt plus3.5pt minus0pt}

\def\small{\@setsize\small{13.6pt}\xipt\@xipt
\abovedisplayskip 13pt plus3pt minus3pt%
\belowdisplayskip \abovedisplayskip
\abovedisplayshortskip \z@ plus3pt%
\belowdisplayshortskip 7pt plus3.5pt minus0pt
\def\@listi{\parsep 4.5pt plus 2pt minus 1pt
     \itemsep \parsep
     \topsep 9pt plus 3pt minus 3pt}}

\relax

\catcode`@=12

\catcode`\@=11

\def\section{\@startsection{section}{1}{\z@}{3.5ex plus 1ex minus
   .2ex}{2.3ex plus .2ex}{\large\bf}}

\def\thesection{\arabic{section}}
\def\thesubsection{\arabic{section}.\arabic{subsection}}

\def\appendix{\setcounter{section}{0}
 \def\thesection{Appendix \Alph{section}}
 \def\thesubsection{\Alph{section}.\arabic{subsection}}
 \def\theequation{\Alph{section}.\arabic{equation}}}

\def\SymBoxes#1#2#3#4{\newdimen\un@t \un@t#3%
\raisebox{#1}{\rule{#2\un@t}{#4}\hskip-#2\un@t
\@tempdimb\un@t \advance\@tempdimb by-#4\@tempcntb#2\relax%
\@whilenum{\@tempcntb>0}\do{
\rule{#4}{\un@t}\hskip\@tempdimb \advance\@tempcntb by\m@ne}%
\hskip-#2\un@t \rule[\un@t]{#2\un@t}{#4}%
\rule[\un@t]{#4}{#4}\hskip-#4
\rule{#4}{\un@t}}\hskip-#4}                

\begin{document}

\newcommand{\beq}{\begin{equation}}
\newcommand{\eeq}{\end{equation}}
\newcommand{\bea}{\begin{eqnarray}}
\newcommand{\eea}{\end{eqnarray}}
\newcommand{\beas}{\begin{eqnarray*}}
\newcommand{\eeas}{\end{eqnarray*}}
\newcommand{\defi}{\stackrel{\rm def}{=}}
\newcommand{\non}{\nonumber}
\newcommand{\bquo}{\begin{quote}}
\newcommand{\enqu}{\end{quote}}
\renewcommand{\(}{\begin{equation}}
\renewcommand{\)}{\end{equation}}
\def\IZ{{\mathbb Z}}
\def\IR{{\mathbb R}}
\def\IC{{\mathbb C}}
\def\IQ{{\mathbb Q}}

\def\CM{{\mathcal{M}}}
\def\dCM{{\left \vert\mathcal{M}\right\vert}}

\def \d{\textrm{d}}
\def \p{\partial}

\def \Pf{\rm Pf\ }

\def \pr{\prime}

\def\Tr{ \hbox{\rm Tr}}
\def\half{\frac{1}{2}}

\def \eqn#1#2{\begin{equation}#2\label{#1}\end{equation}}
\def\de{\partial}
\def\Tr{ \hbox{\rm Tr}}
\def\H{ \hbox{\rm H}}
\def\HE{ \hbox{$\rm H^{even}$}}
\def\HO{ \hbox{$\rm H^{odd}$}}
\def\K{ \hbox{\rm K}}
\def\Im{ \hbox{\rm Im}}
\def\Ker{ \hbox{\rm Ker}}
\def\const{\hbox {\rm const.}}
\def\o{\over}
\def\im{\hbox{\rm Im}}
\def\re{\hbox{\rm Re}}
\def\bra{\langle}\def\ket{\rangle}
\def\Arg{\hbox {\rm Arg}}
\def\Re{\hbox {\rm Re}}
\def\Im{\hbox {\rm Im}}
\def\exo{\hbox {\rm exp}}
\def\diag{\hbox{\rm diag}}
\def\longvert{{\rule[-2mm]{0.1mm}{7mm}}\,}
\def\a{{\textsl a}}
\def\dag{{}^{\dagger}}
\def\tq{{\widetilde q}}
\def\p{{}^{\prime}}
\def\W{W}
\def\N{{\cal N}}
\def\hsp{,\hspace{.7cm}}
\newcommand{\C}{\ensuremath{\mathbb C}}
\newcommand{\Z}{\ensuremath{\mathbb Z}}
\newcommand{\R}{\ensuremath{\mathbb R}}
\newcommand{\rp}{\ensuremath{\mathbb {RP}}}
\newcommand{\cp}{\ensuremath{\mathbb {CP}}}
\newcommand{\vac}{\ensuremath{|0\rangle}}
\newcommand{\vact}{\ensuremath{|00\rangle}}
\newcommand{\oc}{\ensuremath{\overline{c}}}

\def\I{\mathcal{I}}

\def\M{\mathcal{M}}
\def\F{\mathcal{F}}
\def\d{\textrm{d}}

\def\eps{\epsilon}

\begin{flushright}
\end{flushright}

\vspace{20pt}
\begin{center}
{\LARGE {\bf Non-supersymmetric Conifold}}
\end{center}

\vspace{-10pt}
\begin{center}
{\large   {Anatoly Dymarsky$^a$} and  Stanislav Kuperstein$^b$}

\vspace{10pt}

\vspace{10pt}
\textit{\normalsize $^a$ School of Natural Sciences, Institute for Advanced Study,\\Princeton, NJ, 08540\\}
\textsf{dymarsky@ias.edu}
\\

\vspace{10pt}

\textit{\normalsize $^b$ Institut de Physique Th\'eorique, \\
CEA Saclay, CNRS-URA 2306,
91191 Gif sur Yvette, France\\}
\textsf{stanislav.kuperstein@cea.fr}

\end{center}

\vspace{5pt}

\begin{center}
\textbf{\large Abstract}
\end{center}
We find a new family of non-supersymmetric numerical solutions of IIB supergravity which are dual to
the $\mathcal N=1$ cascading ``conifold'' theory perturbed by certain combinations of relevant single trace and marginal double trace operators with non infinitesimal couplings.
The SUSY is broken but the resulting ground states, and their gravity duals, remain  stable, at least perturbatively.
Despite the complicated field theory dynamics the gravity solutions have a simple structure. They feature the Ricci-flat non-K\"ahler metric on the deformed conifold and the imaginary self-dual three-form flux accompanied by a constant dilaton.

\vspace{4pt} {\small

\noindent }

\vspace{1cm}

\renewcommand{\thefootnote}{\arabic{footnote}}
\thispagestyle{empty}
\newpage

\section{Introduction}

The problem of finding a holographic description for QCD is an outstanding challenge. Even in the strongly coupled regime, when the dual background is expected to be weakly curved and potentially can be described by supergravity, to find the corresponding gravity background is beyond our reach. Partially this can be attributed to the lack of supersymmetry. So far supersymmetry was the main vehicle to find new solutions and to guarantee their stability. Because of supersymmetry the dual solutions had a simple and elegant structure, but what we could learn about the dual field theories was bounded to supersymmetric dynamics.

In this paper we want to make a step in the direction of finding a gravity background dual to a non-supersymmetric field theory.  As we are aiming at a confining theory the deformed conifold geometry of the Klebanov-Strassler (KS) background \cite{KS} seems a natural starting point. The KS geometry admits a global $SU(2)\times SU(2)$ symmetry, which, although redundant from the QCD point of view, drastically simplifies the story on the gravity side. We would have to maintain this symmetry throughout  the paper.

At the next step we would like to break supersymmetry by means of perturbing theory by relevant operators, say, by giving mass to gauginos. This will break supersymmetry while preserve the global $SU(2)\times SU(2)$. The corresponding gravity solution is straightforward to find when the gaugino mass is infinitesimally small and SUSY is softly broken \cite{KuSo,KuSo2,aD3}. Conceptually to find the corresponding solutions when the masses are large is also straightforward.  Because the global $SU(2)\times SU(2)$ symmetry is preserved  the corresponding $10d$ IIB supergravity background can be described by the so-called Papadopoulos-Tseytlin (PT) ansatz \cite{PT} that involves only ten functions of radial variable. The supergravity equations of motion reduce to ten coupled non-linear second order ODEs which one would need to solve imposing proper boundary conditions in the IR and the UV.

In practice such a solution can be found only numerically but even in this case it is a formidable task.\footnote{This question is successfully attacked in \cite{Bennett:2011va}.} To make the problem more manageable one needs to employ some trick which would drastically reduce the complexity of the system. For instance this was successfully done in \cite{Butti}  where a new family of gravity solutions dual to the baryonic branch of field theory was found. Thanks to supersymmetry the problem reduced just to two coupled first order ODEs with simple boundary conditions while other eight functions from the PT ansatz are algebraically dependent on the first two.
Later these solutions were generalized by use of a solution generating technique based on string theory dualities \cite{Gaill,Caceres,Eland}. The resulting solutions are still attractively simple but develop various singularities. Typically these solutions have singular metric near the tip.
And whenever solution is IR regular there is a deviation from the asymptotically $AdS$ behavior in UV signaling the presence of the irrelevant dimension eight operator.
These solutions are of no use for us as we aim to find a regular non-supersymmetric background. Moreover we want to preserve the decoupling limit and therefore we do not allow perturbations by irrelevant operators.

A good idea would be to preserve the simplicity of the supersymmetric solution while breaking supersymmetry explicitly.
One example of such a trick in the context of D7-brane embedded into a supersymmetric background was developed in  \cite{DKS7}.
It was shown there that in a special case of constant dilaton background the supersymmetry conditions for D7 can be partially relaxed. Then the supersymmetry will be completely broken but the resulting solution for D7 will maintain the simplicity of a fully supersymmetric one.
It is interesting to note that such a trick fails if the D7-brane is embedded into a more complicated supersymmetric background with running dilaton \cite{D7bb}. In our case  the situation is similar.  Our trick is based on the following observation \cite{GP1,GP2,GKP}. The simplest version of the bulk supersymmetry condition requires the RR four form to be related to the warp-factor $C_4=h^{-1} {\rm Vol}(\R^{3,1})$ and the three-form flux to be a $(2,1)$ primitive form. Furthermore, in the absence of D-branes the dilaton is constant while the unwarped $6d$ metric is K\"ahler and Ricci-flat. A remarkable observation is that if the three-form flux is an arbitrary imaginary self-dual (ISD) form (not necessarily $(2,1)$ primitive) this is enough for the fluxes to decouple from the metric/dilaton equations. Without sources the dilaton will have to be constant and one would end up with a Ricci-flat unwarped $6d$ metric which is not necessarily K\"ahler. A non-K\"ahler metric and a generic ISD flux are both breaking supersymmetry. We will denote this type of solution with the ISD three-form flux, a constant dilaton and a Ricci-flat $6d$ metric as the GKP (for Giddings, Kachru, Polchinski \cite{GKP}) background.  These solutions are significantly simpler than the generic ones as the fluxes completely decouple from the equations for metric.

Relaxing the SUSY condition only for the three-form flux while keeping the metric intact in the context of the KS background does not yield any new non-singular solution \cite{KuSo}. Therefore we are motivated to find a new Ricci-flat, presumably non-K\"ahler, metric on the deformed conifold.
Below we argue that the conventional K\"ahler Ricci-flat metric on the deformed conifod \cite{candelas} -- the unwarped metric of the KS solution -- can be generalized to a one-parametric family of IR-regular Ricci-flat metrics which asymptote to the conventional metric in the UV. These new non-K\"ahler metrics can be used to construct novel gravity backgrounds which are regular everywhere and approach the KS solution in the UV. Therefore these backgrounds must be dual to the original $\mathcal N=1$ cascading ``conifold'' theory perturbed by some particular combinations of SUSY-breaking relevant and/or marginal operators. We do not expect these particular combinations of couplings to be special in any way from the field theory point of view. Quite the opposite, there must be a larger space of SUSY-breaking relevant deformations which lead to well-defined theories with stable vacua.

\begin{figure*}[t]
\centering
 \includegraphics[width=0.6\textwidth]{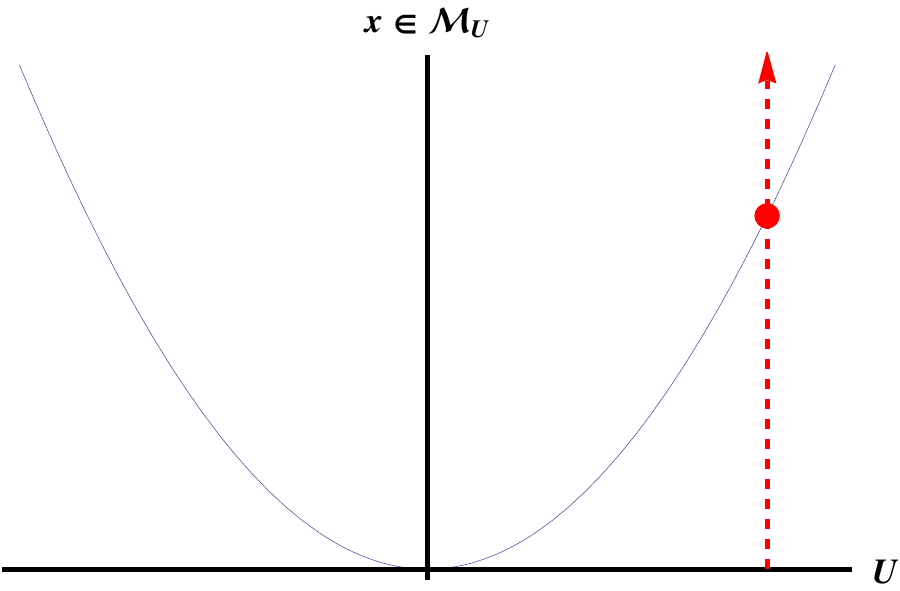}
\caption{We define ${\mathcal M}_U$ as a space of combinations of relevant/marginal couplings -- and the corresponding theories -- such that the bottom component of the $U(1)_{\rm baryon}$ multiplet has the vacuum value $U$. Schematically we denote $x$ to be the coordinates on ${\mathcal M}_U$ (we do not specify the dimension of ${\mathcal M}_U$) such that $x=0$ corresponds to the unperturbed ${\mathcal N=1}$ theory. Red dashed line corresponds to the motion in the space of theories starting from the unperturbed ${\mathcal N=1}$ theory such that vev $U$ stays constant. Most of these theories have complicated gravity dual. At a specific point the gravity dual accidently becomes simpler -- it is of GKP type. The one-dimensional family of such theories forms the thin blue line. These are the theories we study.}
\label{fig:spaceoftheories}
\end{figure*}

To illustrate the point it is convenient to parameterize the family of new backgrounds which we find below, by $U$ -- the vev of the bottom component of the $U(1)_{\rm baryon}$ multiplet. This is because the supersymmetric solution for infinitesimally small $U$ is of the GKP type \cite{GHK1}. Hence $U$ is a good coordinate to parametrize the new one-dimensional family of the GKP backgrounds, at least locally near $U=0$.
Value of $U$ will specify the location of the new non-SUSY vacuum on the baryonic branch of the original SUSY theory. Let us start with the original SUSY theory in a vacuum on the baryonic branch with some non-trivial value of $U$. For finite $U$ the dual gravity background is quite complicated: it is a $SU(3)$-structure solution with running dilaton etc. \cite{Butti}. Then we start turning on SUSY-breaking relevant and marginal couplings such that the vacuum value of $U$ stays intact. Overall there is a large space of such combinations of couplings and the corresponding theories which we denote by ${\mathcal M}_U$. Since SUSY is broken the corresponding gravity backgrounds presumably are even more complicated than the original point of our journey -- the $SU(3)$ structure  solution of \cite{Butti} with the given $U$. Yet, we claim, there is at least one particular point in ${\mathcal M}_U$ such that the gravity dual admits the simple GKP structure -- it has an ISD flux, a constant dilaton and a Ricci-flat metric. These are the theories, and their gravity duals, we study in this paper. The schematic picture is shown in Fig. \ref{fig:spaceoftheories}.

This paper is organized as follows. In the next section we numerically construct a one dimensional family of gravity backgrounds that corresponds to the blue line in fig. \ref{fig:spaceoftheories}. We construct the novel $SU(2)\times SU(2)$ invariant Ricci-flat non-K\"ahler metrics on the deformed conifold and then dress them up by the appropriate fluxes and warping. Then in Section \ref{sec:ft} we discuss stability of the resulting solutions and their meaning from the dual field theory point of view. We outline our results in Section \ref{sec:outline}.

\section{Supergravity solutions}
\label{sec:sugra}

\subsection{Ricci-flat metric on the deformed conifold}
\label{sec:rf}

The starting point of our journey is the observation that the Gubser-Herzog-Klebanov (GHK) solution that corresponds to the vacua of the original ${\mathcal N=1}$ theory with the infinitesimally small $U$ has the GKP structure \cite{GHK1,GHK2}:
\begin{itemize}
  \item  Constant dilaton, $e^\Phi=g_{\rm s}$ ;
  \item  Imaginary self-dual (ISD) three-form flux, $iG_{3} =  \star_6 G_{3}$, where $\star_6$ is the $6d$ Hodge dual and
            $G_{3} \equiv F_{3} + \frac{i}{g_{\rm s}} H_{3}$ ;
  \item  RR four-form $C_4=h^{-1}  {\rm Vol}(\R^{3,1})$, where $h$ is the warp factor:
                  \begin{equation}
                  \ell_s^{-2} d s_{10}^2 = h^{-1/2} d x_\mu d x^\mu + h^{1/2} d s^2_{M_6} \, ;
                  \label{warp}
                  \end{equation}
  \item  Ricci-flat $6d$ unwarped metric on $M_6$.
\end{itemize}
Beyond linear order the baryonic branch solutions of \cite{Butti} do not preserve these  properties. A natural question then is wether it is possible to continue the GHK solution beyond linear order such that the Ricci-flatness of the unwarped metric and other nice GKP properties of the solutions are preserved. We will argue that this is indeed possible and there is a one-dimensional family of such backgrounds which form the blue line in Fig \ref{fig:spaceoftheories} touching the horizontal line, the baryonic branch of the SUSY theory, at the origin.

More broadly we can pose a question of finding non-singular Ricci-flat deformations of the conventional K\"ahler metric on the deformed conifold. As an extra condition we will require the new metric to approach  the original K\"ahler metric in the UV. For simplicity we also preserve the $SU(2)\times SU(2)$ symmetry. The most general metric compatible with these symmetries is given by the PT ansatz  and depends on four functions of the radial variable $\tau$ \cite{PT}\footnote{\label{lambda} It is possible to introduce another fifth function $\lambda(\tau)$ by substituting $\psi\rightarrow \psi +\lambda$ while keeping $d\psi$ intact. Obviously the constant part of $\lambda$ is a pure gauge associated with the action of $U(1)_R$. Introducing $\lambda$ will not lead to any new solutions as the Ricci-flatness requires $\dot{\lambda}=0$.}:
\begin{eqnarray}
\label{metric6d}
\epsilon^{-4/3} d s_{6}^2 &=& \frac{2}{3} e^{-8 p +3 q} \left( d \tau^2 + g_5^2 \right)
       +  e^{2 p +3 q} \Big( \cosh(y) \big(  e^z ( e_1^2+e_2^2 ) + e^{-z} (\epsilon_1^2+\epsilon_2^2) \big)  \nonumber \\
     && \qquad \qquad \qquad \qquad   -2 \sinh(y) \left( e_1 \epsilon_1 + e_2 \epsilon_2 \right) \Big)  \, .
\end{eqnarray}
We refer the reader to \cite{PT} for the definitions of the angular forms $e_i,\epsilon_i,$ and $g_5$. For the KS solution one finds $z(\tau)=0$ and:
\begin{eqnarray}
\label{KSpqy}
    &&
    e^{10 p(\tau)} = K^3(\tau) \sinh(\tau) \, , \quad
    e^{6 q(\tau)}  = \frac{3^{1/2}}{8} K^{4/5}(\tau) \sinh^{8/5} (\tau) \, , \quad
    e^{y(\tau)} = \tanh \left( \frac{\tau}{2} \right) \, , \nonumber \\
    && \qquad \qquad \textrm{where} \quad K(\tau) \equiv \frac{\left( \sinh (2 \tau) -  2 \tau \right)^{1/3}}{2^{1/3} \sinh(\tau)} \, .
\end{eqnarray}
This $6d$ metric is a solution of the $10d$ supergravity EOMs from \cite{PT}, together with zero three-form fluxes and $\Phi={\rm const}$, provided we  redefine some of the metric fields in the following manner (the expression for $A_{\rm PT}$ ensures that the warp factor is one, $h(\tau)=1$)\footnote{Our $z(\tau)$ coincides with the one used in \cite{GHK1} only at the linear order. Also, for $z=0$ our notations reduce to those of \cite{KuSo}. This is the reason for using $p$ in the new notations.}:
\begin{eqnarray}
\label{PTtoHERE}
\left( e^x , e^p, e^{2 A} \right)_{\rm PT}
    &=& \left( 2^{-1/2} 3^{-1/4} e^{2 p + 3 q} , 3^{1/4} e^{p-q}, 2^{-1/2} 3^{-3/4} e^{5 q} \right)_{\rm here} \, , \nonumber \\
e^g &=& \frac{e^z}{\cosh (y)}    \, ,  \quad
a   = e^z \tanh (y)    \, .
\end{eqnarray}
There is a $\Z_2$ symmetry $\I$ that exchanges the two two-spheres of the conifold (and also changes sign of the three-forms). In the coordinates $z_i$ satisfying $\sum_i^4 z_i^2=\epsilon^2$ it acts by $z_4 \rightarrow -z_4$.
This symmetry leaves the variables  $p(\tau)$, $q(\tau)$ and $y(\tau)$ invariant and flips the sign of $z(\tau)$.
In the $z=0$ case the geometry is $\I$ invariant  as in \cite{KS} (see also \cite{KuSo}, \cite{PandoZayas:2001iw}).

For the metric (\ref{metric6d}) to be Ricci-flat the functions $p(\tau)$, $q(\tau)$, $y(\tau)$ and $z(\tau)$ have to satisfy four second order ODEs and a first order constraint (the zero energy condition). Hence the full space of $SU(2)\times SU(2)$ invariant Ricci-flat metrics on conifold is seven-dimensional. Obviously the shift
$\tau \rightarrow \tau + {\rm const}$ is a symmetry which we can fix by choosing the tip of the geometry to be at $ \tau=0$. Another simple parameter is the overall re-scaling of the metric given by $q \rightarrow q + {\rm const}$. Alternatively, this shift controls the size of the three-sphere at the tip of the conifold which can be  identified with the deformation parameter $\epsilon$. Thus we are left with only five non-trivial parameters and $\epsilon$.

The counting above agrees with the linearized analysis in the vicinity of the singular conifold $\sum z_i^2=0$. The warped product of singular conifold and the Minkowski space is a gravity  dual to a particular CFT \cite{KW}. Therefore all small Ricci-flat perturbations of the unwarped metric are in one-to-one correspondence with couplings and vevs of certain operators in field theory. There are three such operators in this case \cite{Ceresole1,Ceresole2,fluxes}:
\begin{enumerate}
  \item The gaugino bilinear $\lambda_1\lambda_1-\lambda_2\lambda_2$ ,
  \item The bottom component of $(W^2\bar{W}^2)_+$ ,
  \item The bottom component of the $U(1)_{\rm baryon}$ multiplet $U\sim {\rm Tr} (|A^2|-|B^2|)$ .
\end{enumerate}
We will denote these operators by their dimensions: 3, 6 and 2 respectively. The first two operators are even,  while the last one is odd under ${\mathcal I}$. Three operators correspond to six small Ricci-flat perturbations of metric on the singular conifold. This is not in contradiction with our previous finding because in the singular conifold case the shift of $\tau = \ln (r^3/\epsilon^2)$ and an overall rescaling of conic metric $d r^2 + r^2 d s^2_{T^{1,1}}$ coincide. Hence we are left with six independent linear modes that correspond to the three operators in the dual field theory.

The analysis of small Ricci-flat perturbations around the deformed conifold should give similar results simply because in the UV region the deformed conifold approaches the singular one, up to  $1/r$ corrections.
Indeed there are six infinitesimal Ricci flat perturbations,  two $\I$ odd and four $\I$ even. The $\I$ odd perturbations correspond to the coupling and the vev of $U$. The four $\I$ even perturbations correspond to the couplings and vevs of the operators of dimension 6 and 3. Because there are two different dimension three operators in this theory -- two combinations of gauginos $\lambda_1\lambda_1\pm \lambda_2\lambda_2$ -- which mix in the deformed conifold case (but not in the singular conifold case) we will not be able to explicitly identify the operator responsible for this metric perturbation. In what follows we will simply denote this operator as $\lambda\lambda$ and call the corresponding coupling the gaugino mass $m_{\lambda\lambda}$. Four $\I$ even modes plus two $\I$ odd modes give six in total, but this is not in contradiction with five non-trivial parameters we found above. This is because one of the six modes is actually the shift of $\tau$ i.e. the $\tau$-derivative of the background solution.

These six infinitesimal Ricci-flat perturbations around the conventional conifold metric have been found explicitly.
Unfortunately the full description is quite technical. We briefly mention it below but a detailed knowledge of these modes is not necessary to follow our logic. An uninterested reader can skip until the next paragraph. The  $\I$-odd Ricci-flat perturbations were found by GHK in \cite{GHK1}. The $\I$-even ones can be found using the formalism to study linear $\I$ even $SU(2)\times SU(2)$ invariant perturbations around the KS solution proposed by Borokhov and Gubser \cite{BG} and later developed in \cite{Bena}. The full space of solutions is parameterized by sixteen constants $X_1,\dots,X_8,Y_1,\dots,Y_8$. To find the Ricci-flat perturbations of the conifold metric one can impose that full background is of the GKP type, i.e. the  dilaton is constant, three-form flux is ISD etc. Then ignoring an overall rescaling one recovers exactly four modes. In the notations of \cite{aD3} these modes are  parameterized by $X_3,X_4,Y_2,Y_3$ while $X_2=-2X_3/3 $, $Y_1=-5Y_2/3$ and all other $X_i$ and $Y_i$ are zero. Only three modes are truly non-trivial as the mode $Y_2={\rm const}, Y_3=X_3=X_4=0$ is the derivative of the background with respect to $\tau$. It can be attributed to the vev of an operator of dimension 3 in a sense that the linear mode asymptotes to $1/r^3$ at infinity. Then $X_4$ corresponds to the coupling $m_{\lambda\lambda}$ while $X_3,Y_3$ correspond to the coupling and the vev of the dimension six operator correspondingly.

Since we know the linear Ricci-flat perturbations explicitly we can analyze their behavior in the UV and IR. There is only one UV divergent mode which corresponds to turning on the coupling of the irrelevant dimension six operator in the dual field theory. To keep the UV asymptotic the same as in the conventional case and hence the field theory well defined in the UV we want to keep this mode ``turned off''. In the IR there are two regular modes. One is  $\I$ odd and corresponds to the infinitesimal motion along the baryonic branch. The other is the $\I$ even mode that turns on the dimension six coupling -- the same mode we discussed above.
So there is no Ricci-flat IR and UV regular $\I$ invariant infinitesimal deformation of the conventional deformed conifold metric. To find a new Ricci-flat background one has to consider the $\I$ breaking modes as well.

The two IR regular linearized modes can be extended  into two-parameter family of IR regular Ricci-flat metrics on the conifold.
With vanishing fluxes, constant dilaton and $h=1$ the IIB supergravity action of \cite{PT} reduces to the $6d$ Hilbert-Einstein action, and so we can read from it the four Ricci flatness equations.  We relegated these equations to Appendix A.
The two-parameter family of the IR regular solutions is given\footnote{We choose the field parametrization here ($e^{10p(\tau)}$ instead of $p(\tau)$, $\dot{q}^{-1}(\tau)$ instead of $\dot{q}(\tau)$ etc) so that all the expressions will be regular both at the origin and for large $\tau$.} by:
\begin{eqnarray}
\label{IRbc}
 e^{10 p (\tau)} &=& \frac{2}{3} \tau   + \left(  - \frac{4}{45} + \zeta_2 \right) \tau^3 + \ldots \, , \nonumber \\
 \frac{1}{\dot{q} (\tau)} &=& \frac{15}{4} \tau   + \left( - \frac{7}{8} + \frac{45}{26} \cdot \zeta_2 \right) \tau^3 + \ldots \, , \nonumber \\
 e^{y (\tau)} &=&
     \frac{1}{2} \tau   + \left( - \frac{1}{24} - \frac{15}{52} \cdot \zeta_2 - \frac{1}{2} \cdot \zeta_1^2\right) \tau^3 + \ldots \, , \nonumber \\
 z (\tau) &=&
   \zeta_1 \tau^2   + \left( - \frac{7}{15} \cdot \zeta_1 - \frac{4}{5} \cdot \zeta_1^3 - \frac{15}{13} \cdot \zeta_2 \zeta_1 \right) \tau^3 + \ldots \, .
\end{eqnarray}
We write here only the result for $\dot{q} (\tau)$ since adding constant to $q(\tau)$ corresponds to the overall rescaling of the metric (\ref{metric6d}). This rescaling can be in turn absorbed in a redefinition of the deformation parameter $\epsilon$. Fixing of $\epsilon$ in the new solution is an important issue and we will specifically address it in the end of this section.

We introduced here two parameters $\zeta_1$ and $\zeta_2$ such that the $\I$ symmetry flips the sign of $\zeta_1$ while keeping $\zeta_2$  invariant. The point $(\zeta_1, \zeta_2) = (0,0)$ is the conventional deformed conifold metric, i.e. the unwarped metric of the KS solution, while the infinitesimal $\zeta_2$ and $\zeta_1$ correspond to ``turning on'' the dimension six mode and the GHK mode respectively. Notice that both $\zeta_1$ and $\zeta_2$ appear in the sub-leading order in $p(\tau)$, $q(\tau)$ and $y(\tau)$ and so the solution is indeed regular at $\tau=0$ for any $\zeta_2$ and $\zeta_1$ exactly as the original KS background.

Let us stress again that we solve the full-nonlinear Ricci-flatness equations and $\zeta_2$, $\zeta_1$ are \emph{not} assumed to be infinitesimally small in \eqref{IRbc}.

\begin{figure*}[tb]
\centering
  \includegraphics[trim = 0mm 40mm 0mm 70mm, width=0.6 \textwidth]{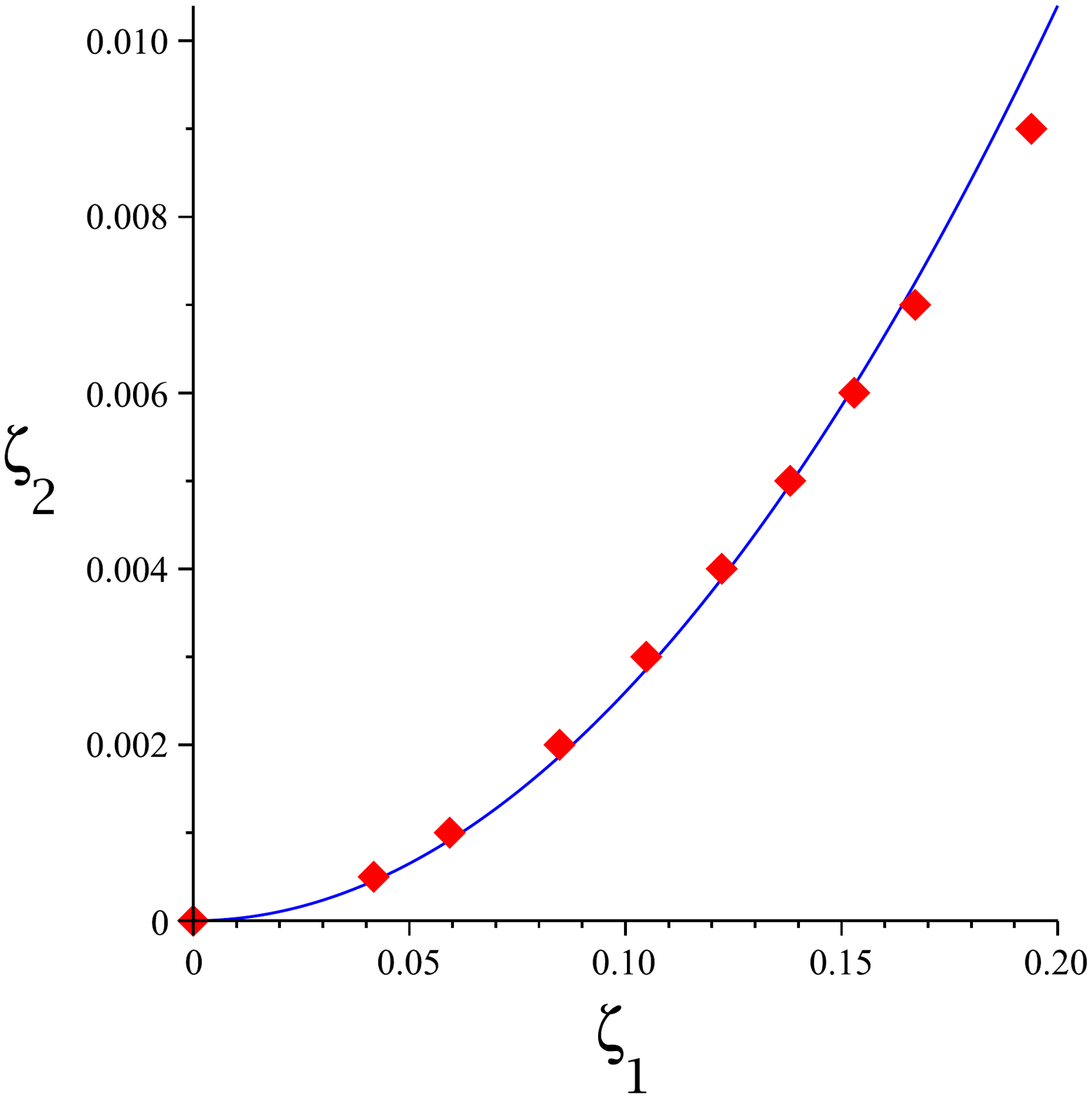}
\caption{The red points show the values of $\zeta_1$ and $\zeta_2$ for which the Ricci-flat metric is regular in both the IR and the UV. Notice that the points
near the origin lie on the parabola (the blue line). The deviation  of the points with larger $\zeta_1$ from the parabola most likely indicates the contribution of the $\zeta_1^4$ term in (\ref{familyc}).}
\label{fig:c1c2}
\end{figure*}

Regularity in the UV provides only one constraint on $\zeta_{1,2}$ and therefore we expect to find a one-parameter family of non-singular Ricci-flat metrics on the deformed conifold, which approach the conventional metric at large radius.
The boundary conditions at $\tau = \infty$ should coincide with the deformed conifold values (\ref{KSpqy}):
\begin{equation}
\label{KSasympt}
 e^{10 p (\tau)} \approx 1 \, , \quad  \frac{1}{\dot{q} (\tau)} \approx \frac{9}{2} \, , \quad
 e^{y (\tau)} \approx 1 \, , \quad z (\tau) \approx 0 \, .
\end{equation}
In Appendix A we study the subleading terms and show that the modes of dimension $2$, $3$ and $6$ indeed appear in the large $\tau$ expansion as expected.

\begin{figure*}[t]
\centering
\includegraphics[trim = 0mm 40mm 0mm 70mm, width=0.5 \textwidth]{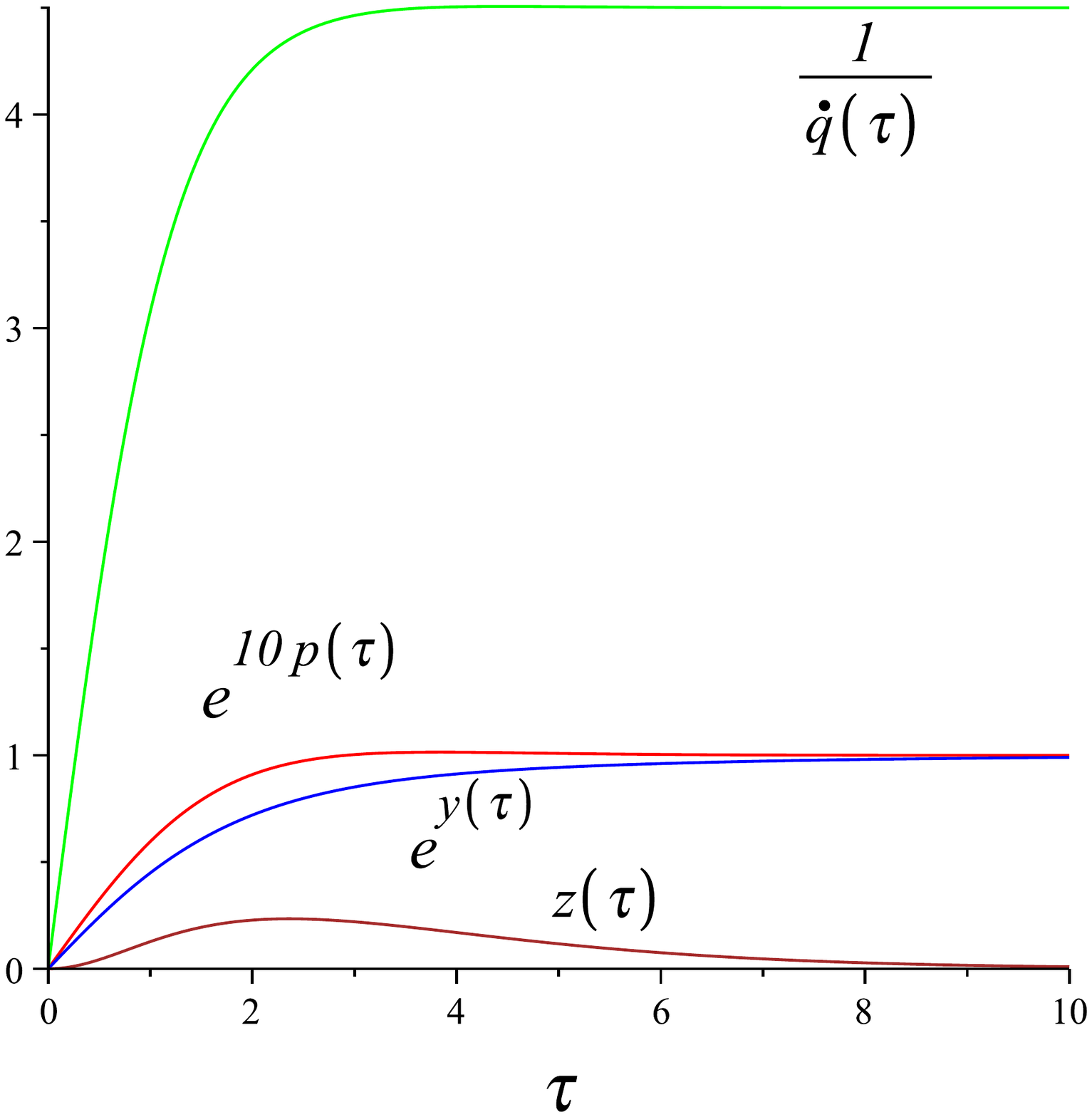}
\caption{The four lines show numerical solutions for $e^{10 p (\tau)}, 1/\dot{q}(\tau), e^{y (\tau)}$ and $z(\tau)$ for $(\zeta_1, \zeta_2) = (0.193(9), 0.009)$ (the last point on the previous plot).}
\label{fig:pqyz}
\end{figure*}

\begin{figure*}[t]
\centering
\includegraphics[trim = 0mm 40mm 0mm 70mm, width=0.5 \textwidth]{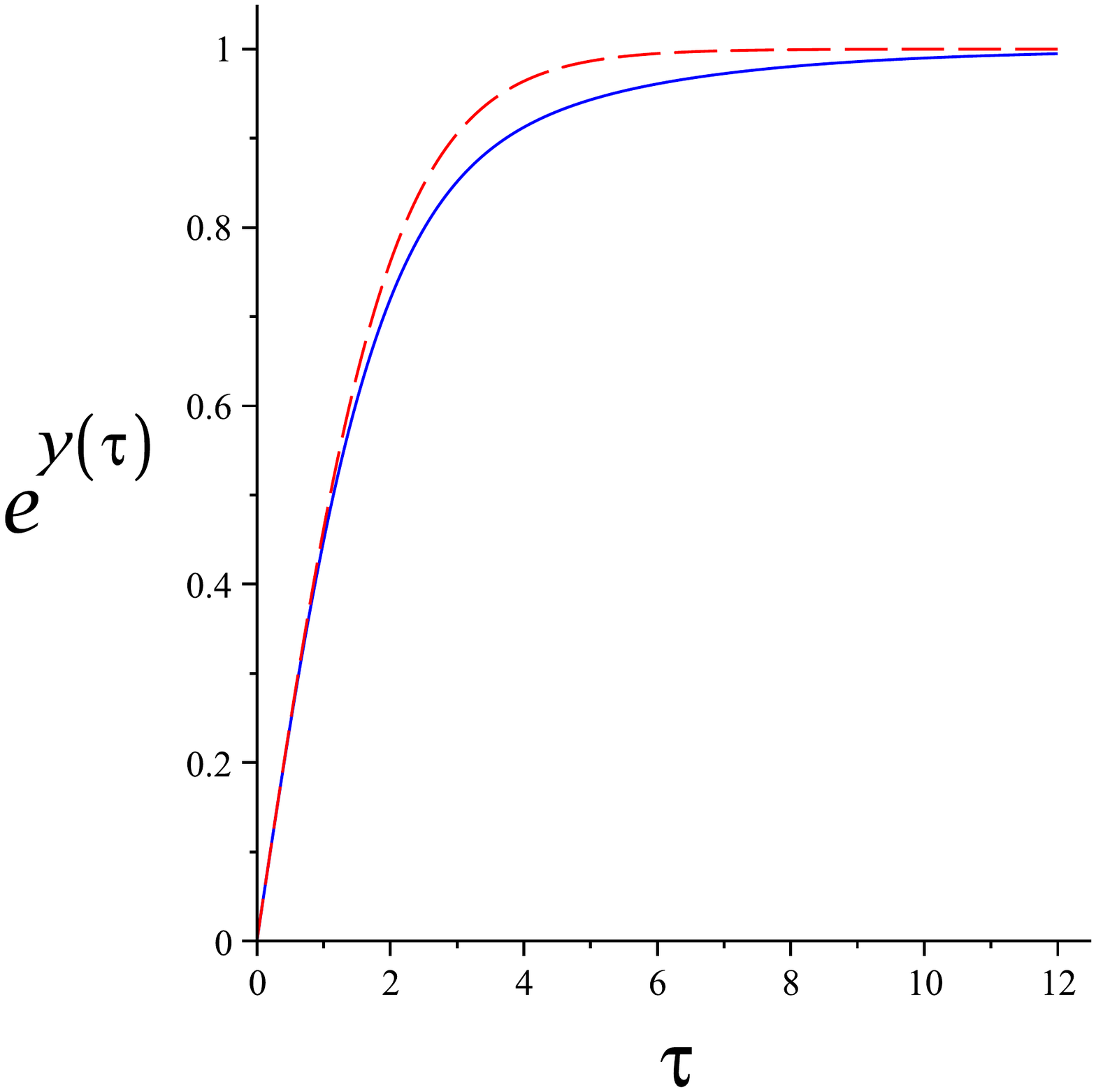}
\caption{The blue solid line show the numerical result for $e^{y (\tau)}$ with $(\zeta_1, \zeta_2) = (0.193(9), 0.009)$, while the red dashed line corresponds to $e^{y_0 (\tau)} = \tanh(\frac{\tau}{2})$ of the conventional deformed conifold of \cite{candelas}. Notice that red line converges to the asymptotic faster than the blue line. This is so because for large $\tau$ the function $1-e^{y(\tau)}$ goes like $e^{-\tau}$ for the the deformed conifold, and as $e^{-\tau/3}$ for our solution.}
\label{fig:exp(y)}
\end{figure*}

Near the origin the leading perturbation (see the $z(\tau)$ expansion in (\ref{KSpqy})) is proportional to $\zeta_1$ (meaning it is $\I$ odd) and therefore at least for small $\zeta_2$ and $\zeta_1$ is it convenient to parameterize the family of new Ricci-flat metrics by $\zeta_1$:
\begin{eqnarray}
\label{familyc}
\zeta_2 ( \zeta_1) \sim \zeta_1^2 + \mathcal{O}( \zeta_1^4) \, ,
\end{eqnarray}
where the $\zeta_1^3$ term is ruled out because of the $\I$-parity.
We found the function $\zeta_2 = \zeta_2 (\zeta_1)$ numerically using the shooting technique and confirm a generic prediction that the expansion (\ref{familyc}) starts with $\zeta_1^2$ (see Figure \ref{fig:c1c2}).
We also present a numerical solution for the functions $e^{10 p (\tau)}, 1/\dot{q}(\tau), e^{y (\tau)}, z(\tau)$ for some particular values of  $\zeta_1$ and $\zeta_2$, see Figure \ref{fig:pqyz}. Finally, on Figure \ref{fig:exp(y)} we compare our numerical results for $e^{y (\tau)}$ with its analytical conventional (KS) counterpart of \cite{candelas}.

It would be interesting to know how far one can move along this new family of solutions and if there is any limit on the resulting value of $U$. These questions would require careful numerical studies which we leave for the future.

\subsubsection*{Ricci-flat metric on deformed conifold: a summary}
The construction of the novel family of Ricci-flat metrics on deformed conifold presented above constitutes a mathematical result which could be of interest in its own right, with no connection 
to the ten dimensional supergravity or dual gauge theory. To make this aspect of our work more accessible to  reader less familiar with the holographic context we briefly outline here main steps
of section \ref{sec:rf}. Our goal was to construct novel Ricci-flat metrics on the deformed conifold -- the complex three-fold described by the equation $\sum_i^4 z_i^2=\epsilon^2$. We construct the Ricci-flat metric by help of the ansatz 
(\ref{metric6d}). This ansatz is by no means general. In particular it explicitly preserves the $SU(2)\times SU(2)$ symmetry. Since all functions $p,q,y,z$ depend only on the radial coordinate $\tau$ the Ricci-flatness condition $R_{\mu\nu}=0$ reduces to a set of coupled ordinary differential equations (\ref{EOMpqyz}). Our ansatz (\ref{metric6d}) is invariant under the shift $\tau\rightarrow \tau+{\rm const}$. We choose this freedom to set $\tau=0$ as the origin
--  the tip of he conifold. That would be reflected in the appropriate behavior of the metric there i.e. particular boundary conditions for $p,q,y,z$ at $\tau=0$. 
We find a two-parameter family of solutions regular at $\tau=0$, i.e. satisfying these boundary conditions, by expanding in power series in $\tau$ for small $\tau$. The result is given by \eqref{IRbc}. Two integration constants $\zeta_{1,2}$ are free parameters at this point. We want the metric to approach the usual conical form at large radius $\tau\rightarrow \infty$. Here by usual we mean the canonical form of the metric on the cone over $T^{1,1}$. There is one Ricci-flat metric on the deformed conifold, found by Candelas and de la Ossa \cite{candelas},  which does just that. This particular metric is captured by our ansatz. At the next step we study the behavior of linear fluctuations around this metric, all within our ansatz of choice. The system of linear equations reveals that there is only one mode which tends to grow at large $\tau$. All other modes actually vanish in this limit. On general grounds we expect that by choosing appropriate infinitesimal $\zeta_1,\zeta_2$ one can fine-tune the behavior of the metric near $\tau\rightarrow 0$ such that the growing linear mode will not be present at large $\tau$. Hence the behavior at large $\tau$ would be the same as in the solution of  \cite{candelas}, as well as the linear analysis around it: there is still only one growing mode at large $\tau$. In this way we find that there must be a line in the $\zeta_1,\zeta_2$ plane such that for each point on this line the resulting metric has the same asymptotic behavior at large $\tau$. These metrics are Ricci-flat by construction, as well as regular everywhere including appropriate behavior at the tip $\tau=0$ and at infinity $\tau\rightarrow \infty$.

\subsection{Adding fluxes and warping}

In the previous subsection we outlined the way to find a regular Ricci-flat metric on the deformed conifold such that it approaches the conventional metric at large radius. The next step is to supplement such a metric with a smooth ISD three-form flux which would approach the KS asymptotic in the UV. Below we give an argument why this is possible for any metric found in the previous section.

We start with the PT ansatz for the three-form flux which automatically includes $M$ units of RR flux through the three sphere at the tip and write down the ISD condition:
\begin{eqnarray}
\label{fkFchi}
\dot{f} &=& ( \cosh(y) \cosh(z) + \sinh(y) )^2 - F \cdot \big( \cosh(2 y) + \sinh(2 y)  \cosh(z) \big)     \nonumber \\
\dot{k} &=& \cosh^2(y) \sinh^2(z) + F \cdot \big( \cosh(2 y) - \sinh(2 y)  \cosh(z) \big)     \nonumber \\
\dot{F} &=& \frac{1}{2} (k - f)      \\
\dot{\chi} &=& - \cosh(y) \sinh(z) \Big( ( \cosh(y) \cosh(z) + \sinh(y) ) - 2 F \cdot \sinh(y)  \Big)   \, .  \nonumber
\end{eqnarray}
These functions are related to the functions used in \cite{PT} in the following way:
\begin{equation}
\left( h_1, h_2, b, \chi \right)_{\rm PT} = \left( -P g_s (k+f), -P g_s (k-f), 2 F - 1, 2 P \chi \right)_{\rm here} \, ,
\end{equation}
where $P$ is a constant proportional to $M$, $P = - M \ell_s^2/4$. It is easy to verify that for $e^{y(\tau)}=\tanh(\tau/2)$ and $z(\tau)=0$ this system of equations reduces to the one discussed originally in \cite{KS}.

Na\"ively the solutions are parameterized by four integration constants. One of the constants is a shift of
$\chi\rightarrow \chi + {\rm const}$ which corresponds to a gauge transformation
$B \rightarrow B + {\rm const}\cdot dg_5$ and hence is unphysical. Another constant is a shift $(f,k) \to (f+b,k+b)$
for a constant $b$ which corresponds to a ``large gauge transformation'' $B \rightarrow B + b \cdot w_2$, where $w_2$
is a Betti form on the base of the conifold. This shift
introduces extra Page D3-brane charge to the system. This symmetry will be fixed by the requirement that the Page D3-charge vanishes at the origin, namely that $B\wedge F_3=0$ at $\tau=0$. Therefore there are only two non-trivial parameters left.

In the UV the metric and fluxes should approach the KS values. Linear analysis around the KS background shows that apart from
the simultaneous shift of $f$ and $k$ there are indeed two modes \cite{KuSo}: one is UV regular and the other is UV singular diverging as $e^\tau$. The former is related to the $\Delta=3$ operator, while the latter is the $(0,3)$ flux dual to the $\Delta=7$ operator $\int d^2\theta (W^2\bar{W}^2)_+$. Eventually we would like to ``turn off'' this irrelevant operator ``killing'' one of the two free parameters.

\begin{figure*}[tb]
\centering
  \includegraphics[trim = 0mm 40mm 0mm 70mm, width=0.6 \textwidth]{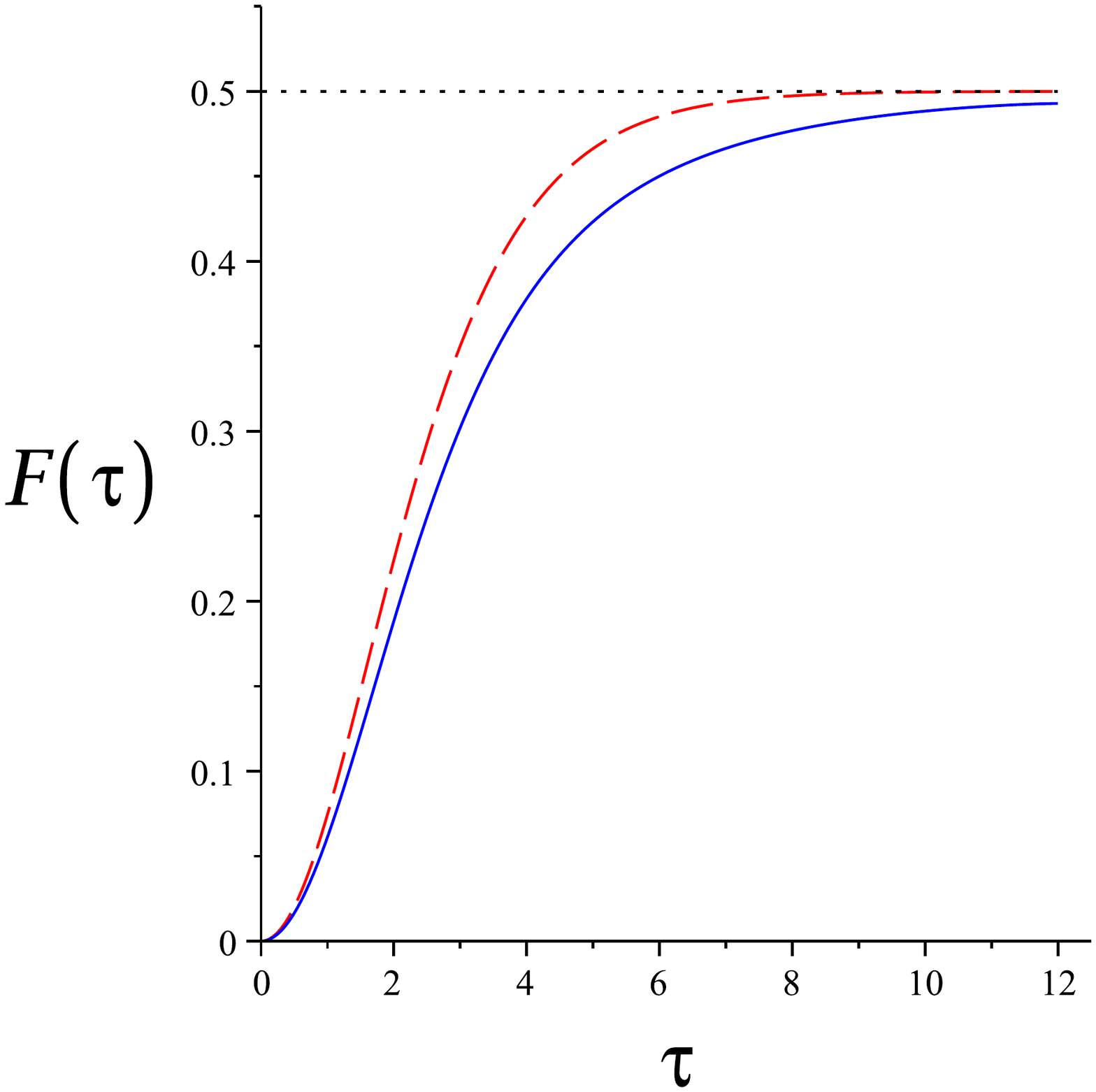}
\caption{The solid blue and the dashed red lines correspond to the numeric solution for $F(\tau)$ with $\zeta_1 = 0.193(9)$ and the same function in the KS geometry, respectively.}
\label{fig:F}
\end{figure*}

It seems like we get a one-parameter family of UV-regular solutions of the ISD equations (\ref{fkFchi}). This is not the end of the story, however, as we have yet to address the IR regularity. Differentiating $f(\tau)$ and $k(\tau)$ we can arrive at a second order PDE for $F(\tau)$ only. Using (\ref{IRbc}) we see that, exactly as in the KS case, near $\tau=0$ the two solutions of this equation are $\tau^{-1}$ and $\tau^2$. Plugging the latter into the equations for $f(\tau)$ and $k(\tau)$  one finds that these functions have the same regular behavior in the IR like in \cite{KS}. This in turn implies that $|F_3^2|=|H_3^2|$ does not diverge at $\tau=0$ and so the background is IR regular.
To be more specific, requiring that $f(\tau)$ and $k(\tau)$ vanish at $\tau=0$ we derive from (\ref{fkFchi}) and (\ref{IRbc}) that there is a one dimensional family of IR regular ISD fluxes parametrized by $\zeta_3$:
\begin{equation}
 F(\tau) \approx  \left( 1 + \zeta_3 \right) \cdot \frac{\tau^2}{12} \, , \quad
 f(\tau) \approx   \frac{\tau^3}{12} \, , \quad
 k(\tau) \approx  \left( 1 + \zeta_3 \right) \cdot \frac{\tau}{3} \, , \quad
 \dot{\chi} \approx - \frac{1}{6} \zeta_1 \left( 4 + \zeta_3 \right) \cdot \tau^2   \, .
\end{equation}
In particular $\zeta_1$, $\zeta_3 = 0$ for the KS background.

\begin{figure*}[tb]
\centering
  \includegraphics[trim = 0mm 40mm 0mm 70mm, width=0.6 \textwidth]{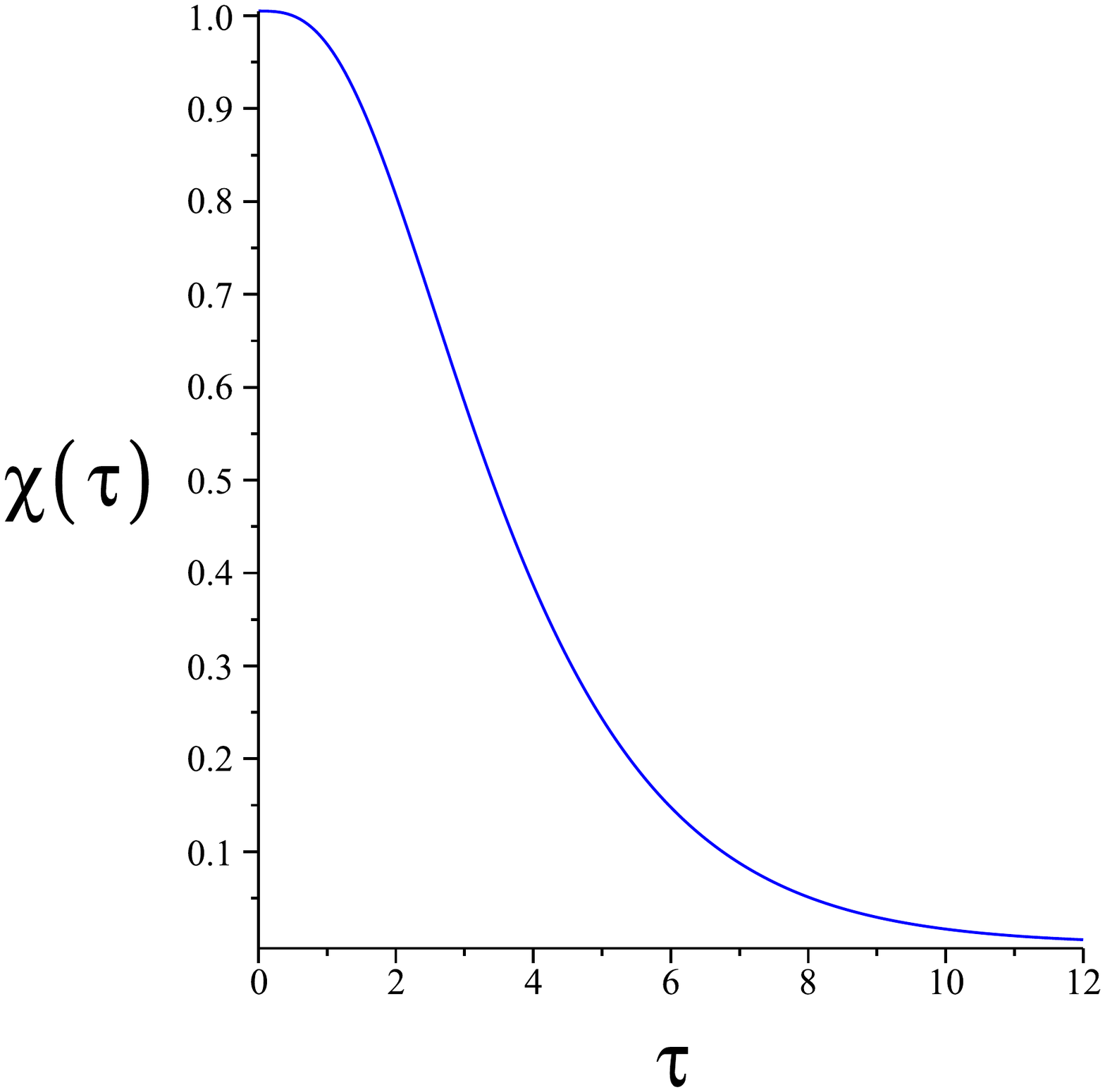}
\caption{The numerical solution for $\chi(\tau)$ with $\zeta_1 = 0.193(9)$.}
\label{fig:chi}
\end{figure*}

We now have to find $\zeta_3$ so that the divergent $\Delta=7$ mode is ``turned off" in the UV. Then for large $\tau$ we have  $F(\tau) \approx 1/2$, while both $f(\tau)$ and $k(\tau)$ go like $\tau/2$, reproducing the UV behavior of the KS solution.  Employing the shooting technique once again we found that $\zeta_3=0.186(8)$ for $\zeta_1 = 0.193(9)$
(the last point in Figure \ref{fig:c1c2}). The corresponding numeric solution for $F(\tau)$ is shown in Figure \ref{fig:F}.
With $F(\tau)$ at hand, we can easily calculate the remaining three functions.
In Figure \ref{fig:chi} we provide a plot for $\chi(\tau)$. As a matter of convenience, we fixed the integration constant so that $\chi(\tau)$ vanishes for large $\tau$, see Figure \ref{fig:chi}.

To complete our discussion of the ten dimensional supergravity background we need to accompany our solution by the warp factor
of the $10d$ metric (\ref{warp}). The equation for $h(\tau)$:
\begin{equation}
 h(\tau)=  \frac{\sqrt{3}}{8} \alpha \int_{\tau}^\infty d x \, e^{-4 p(x) -6 q(x)} \big( (1-F(x)) f(x) + F(x) k(x) \big) \, ,
\end{equation}
where:
\begin{equation}
\alpha \equiv 4 \left( g_s M \ell_s^2 \right)^2 \cdot \epsilon^{-8/3} \, .
\end{equation}
This equation immediately reveals that $h(\tau)$ is regular\footnote{We fixed the integration constant in the equation for $h(\tau)$ by requiring that the warp function vanishes at infinity exactly like in the unperturbed KS background.} at the tip and approaches the KS behavior at large $\tau$.

The last touch is to fix the deformation parameter $\epsilon$ (as we have already mentioned below (\ref{IRbc}) this is equivalent to adding a constant to the field $q(\tau)$).

Since $\epsilon$ is the only dimensional parameter of the solution we can only talk about fixing it
in the context of comparing $\epsilon$ for two different solutions. We want to fix $\epsilon(\zeta_1)$ such that
the corresponding family of gravity backgrounds would correspond to the same dual field theory perturbed by some relevant operators in the UV. In particular, this would require running of the Maxwell D3-charge\footnote{To be more precise, one has to compare the running of the Page charges, since unlike the Maxwell charge it is quantized and so has the proper interpretation as the gravity dual of the cascading gauge theory rank. It turns out, however, that the matching of the Maxwell charges yields exactly the same result for $\epsilon$.} $Q_{D3}$ with the scale to be the same at large radius $r$ for all $\zeta_1$. A convenient reference point is $\zeta_1=0$, i.e. the KS solution.
Thus, to match the KS theory in the UV we require:
\bea
\label{d3cond}
Q_{D3}(r_c) = Q^{KS}_{D3}(r_c)
\eea
at a sufficiently large cutoff radius $r_c$.
Here the Maxwell D3-charge is (see, for example, \cite{aD3}):
\bea
Q_{D3}(\tau) = \frac{g_s}{\pi} M^2 ( (1-F(\tau)) f(\tau) + F(\tau) k(\tau)) \, .
\eea
It seems that this expression is completely independent of $\epsilon$. Notice, however, that $\tau=\ln(r^3/\epsilon^2)$ and so requiring that the two Maxwell charges in (\ref{d3cond}) coincide for the same $r=r_c$ (but different $\tau$'s) we can read from (\ref{d3cond}) the ratio between the KS parameter $\epsilon_{KS}$ and the parameter of the perturbed theory, $\epsilon$.
For instance, for the theory corresponding to the last point on Figure \ref{fig:c1c2} (the one with $\zeta_1= 0.193(9)$) we have $Q_{D3}(\tau=11.866(2))=Q_{D3}^{KS}(\tau=12)$ which  implies $\epsilon/\epsilon_{KS}= 1.069$. It is worth emphasizing here that for large enough $r_c$ (or equivalently $\tau$) the value of $\epsilon/\epsilon_{KS}$ is not sensitive to the variation of the cut-off, since asymptotically $Q_{D3}(\tau) \sim \tau + {\rm const}$.

A more detailed discussion of the condition (\ref{d3cond}) and the resulting constraint on $\epsilon$
in the context of SUSY and non-SUSY vacua of the KS theory can be found in \cite{aD3}.

\section{Dual field theory and stability}
\label{sec:ft}

In this section we discuss the field theories dual to our gravity backgrounds. We suggest that these non-supersymmetric theories correspond to the KS theory perturbed by both a single and a double-trace operators. We also argue that, at least for large $M$, these theories are long-lived both perturbatively and non-perturbatively.

\subsection{The dual field theory interpretation}

Once we outlined the way to find the gravity backgrounds we want to analyze their meaning from the dual field theory point of view. As we require the backgrounds to approach the KS solution in the UV the
resulting field theory must be the original cascading $\mathcal N=1$ ``conifold'' theory perturbed by some relevant and/or marginal operators. In the single trace sector these are the $\I$-odd bottom component of the $U(1)_{\rm baryon}$ multiplet $U$ and the gaugino bilinears. It is easy to see that the marginal top components of $W^2_{\pm}$ are not turned on: the value of dilaton and running of the Maxwell D3-charge are the same for large $r$ as in the KS case.  (In any way adding top components of $W^2_{\pm}$ would only result in the renormalization of gauge couplings).  Hence the solutions in question are dual to the vacua described by the $\N=1$ KS theory perturbed by some, not necessary small, combination of the gaugino masses $m_{\lambda_1\lambda_1}\pm m_{\lambda_2\lambda_2}$
and the operator $U$.
We will not be able to identify the particular combination of $m_{\lambda_1\lambda_1}\pm m_{\lambda_2\lambda_2}$ that is getting turned on (see our discussion in the previous section) and in what follows will refer to this combinations simply as $m_{\lambda\lambda}$.
The vacuum solutions also acquire a non-trivial vev of $U$, as well as vevs of other relevant and irrelevant operators.

For most operators it is straightforward to distinguish the vev from the coupling as the two modes have different powers of $1/r$ where the radial coordinate
\begin{eqnarray}
r=\epsilon^{2/3}e^{\tau/3}\ .
\end{eqnarray}
The operator $U$, however, has dimension two and therefore the two modes mix. The corresponding wave-function, which for large radius can be defined through:
\begin{eqnarray}
\delta ds^2= \sqrt{h(r)} (e_1^2+e_2^2-\epsilon_1^2-\epsilon_2^2)\Psi_U
\end{eqnarray}
has the following asymptotic \cite{GHK1}:
\begin{eqnarray}
\label{psiu}
\Psi_U=\beta \ln \left( \frac{r}{r_0} \right) + \alpha\ .
\end{eqnarray}
Here $r_0$ is some IR scale which we choose such that, in the absence of multi-trace deformations, the coefficient $\beta$ corresponds to the vev of $U$ while $\alpha$ corresponds to its coupling.
In general $r_0$ may depend on various couplings of the theory, such as $m_{\lambda\lambda}$. In the KS case, when all relevant couplings (including $m_{\lambda\lambda}$), are turned off, $r_0=\epsilon^{2/3}e^{1/3}$ \cite{GHK1}.

We do not know $r_0(m_{\lambda\lambda})$, although this can be established in principle. Therefore even if we find our solutions numerically with a very high precision, we will not know $\alpha(\zeta_1)$.\footnote{In practice the main difficulty here is to find the solutions numerically with enough precision to distinguish small $\alpha$ which appears only in the $\zeta_1^3$ order from the large logarithmic contribution $\beta\ln (r/r_0) \sim \zeta_1 \tau $. }  Nevertheless, on general grounds we expect:
\bea
\label{alphabetascaling}
\alpha\sim \zeta_1^3+ {\mathcal O}(\zeta_1^5) \, , \qquad
\beta\sim \zeta_1+ {\mathcal O}(\zeta_1^3) \, .
\eea
This scaling is easy to explain. $\beta$ corresponds to the vev of $U$ which, by definition of $\zeta_1$, scales as $\zeta_1$. Moreover at the linear order in $\zeta_1$ the solution coincides with the GHK background, i.e. has zero coupling $\alpha$. Since $\alpha$ is odd under $\I$ symmetry, as well as $\zeta_1$, it can only start with a cubic term. Let us demonstrate that the same scaling is compatible with (\ref{psiu}). From the fact that $m_{\lambda\lambda}$ is $\I$-even we find that:
\begin{eqnarray}
m_{\lambda\lambda}\sim \zeta_1^2+ {\mathcal O}(\zeta_1^4)\ .
\end{eqnarray}
Now, we generally expect $r_0$ to be $m_{\lambda\lambda}$-dependent such that $\partial r_0/ \partial m_{\lambda\lambda}\neq 0$.
The change in $r_0$ can be reabsorbed into $\alpha$ which indeed indicates that $\alpha\sim \zeta_1^3$.

A non-zero value of $\alpha$ implies that besides the gaugino mass  the dual field theory is perturbed by the dimension two operator $U$. Definitely this can not be the full story: the theory perturbed by the potential $V=aU$ does not have a vacuum at small $U$. We find a consistent interpretation only at the next order in $U$ -- at least for small $\zeta_1$ the gravity solutions in question can be interpreted as a dual to the theory perturbed by the  marginal double-trace\footnote{The dimension of the double trace operator $U^2$ is twice the dimension of $U$ up to  $1/N$ corrections.}
operator $c U^2/2$. The coupling $c$ will alter the boundary conditions in the UV and hence the dual field theory interpretation of $\alpha$. More precisely, when the field theory perturbed by a potential:
\begin{eqnarray}
V = a U + c \frac{U^2}{2} \, ,
\label{V}
\end{eqnarray}
the boundary condition for $\Psi_U$ becomes \cite{Witten, Berkooz:2002ug}:
\begin{eqnarray}
\alpha = a + c\beta \, .
\label{alpha}
\end{eqnarray}
When $c = 0$ we get the usual relation $\alpha = a$, i.e. in this case $\alpha$ is the coupling of $U$.
Since the theory with the $V=aU$ potential has no vacuum state there should be no IR regular solution with $\alpha=a\neq 0$.
In the general $c \neq 0$ case the resulting field theory has a unique vacuum at $U = -a/c$.
The corresponding gravity dual background should have $\alpha$ and $\beta$ such that they satisfy (\ref{alpha}) and the geometry is  regular in the IR. Since the IR regularity requires $\alpha=0$, we see that $\beta=-a/c$, in full agreement with $\beta=U$ and $dV/dU=0$.

It is important to note that the gravity solution itself can not distinguish between different combinations of $a,c$ which result in the same vev of $U$.\footnote{We thank I. Klebanov and J. Maldacena for discussing this point.} In fact the same gravity solution admits interpretation as both being dual to a field theory with and without the double trace deformation (see \cite{Hertog:2004rz,Hertog:2005hu} and  \cite{Maldacena:2010un} correspondingly).
Using this freedom we would like to interpret our solutions as the gravity dual of the field theory perturbed by certain single trace relevant operators and also the marginal double-trace operator $U^2$.
Our logic here was to show that it is a consistent interpretation when in the dual field theory $a \sim \zeta_1$ and $c \sim \zeta_1^2$. 

Strictly speaking, the consideration above and the equivalence between vacuum value of $U$ in the field theory and in the bulk is only valid as long as the theory with $a=c=0$ has a flat direction for VEV of $U$. This is indeed true for infinitesimal $\zeta_1$ as it it can be shown (using the charges under R-symmetry) that the gaugino masses can only contribute to the potential for $U$ at the subleading order $m_{\lambda\lambda}^2\sim \zeta_1^4$.
But we believe that our interpretation -- that the solutions we found are dual to the KS theory perturbed by gaugino masses and the potential $V=aU+cU^2/2$ --  should be correct beyond the infinitesimal order in $\zeta_1$.
In this case the gaugino masses induce a potential for $U$ (on top of $V=aU+cU^2/2$ which we add ``by hands'' in the UV) and the calculation of vacuum value of $U$ is obscured on the field theory side. Similarly on the supergravity side (\ref{alpha}) still holds but we do not know which combination of $\alpha,\beta$ corresponds to the IR regular background. Thus we can not easily compare field theory and gravity to prove our point. At the same time the fact that we included all relevant/marginal operators (allowed by symmetries) into consideration and the fact that this interpretation works for small $\zeta_1$, suggests that the same interpretation should be correct beyond the infinitesimal order in $\zeta_1$.

\subsection{Stability}

The field theory interpretation suggested above ensures  the perturbative stability for infinitesimally small $\zeta_1$ as long as $c>0$ (see (\ref{V})). There are many field theories dual to the same gravity background  and we choose the one with positive $c$: for any theory with $a,c$ dual to a given background the theory with  $-a,-c$ is dual to the same background as well. Indeed when $\alpha=0$ this flipping will not modify $\beta$, as one can see from (\ref{alpha}), and so the gravity solution will be the same.

On the gravity side the stability would follow from the analysis of small perturbations around the classical solutions in question with the boundary conditions:
\begin{eqnarray}
\delta \alpha =c\ \delta \beta\ .
\end{eqnarray}
Clearly the sign of the perturbation mass squared will be sensitive to the sign of $c$. More specifically, in solving the equations for massive modes, the sign of the $4d$ dimensional mass squared will be sensitive to the relative sign of $\delta \alpha$ and $\delta \beta$, which in turn is fixed by $c$.

By interpreting the gravity solutions as being dual to the field theory perturbed by the double-trace operator we ensured perturbative stability, at least for the solutions with infinitesimally small $\zeta_1$.
For larger $\zeta_1$ we would need to calculate the mass-squre $m^2_U$ of small perturbations of $U$ to check that is it positive. For infinitesimally small $\zeta_1$, $m^2_U\sim c$ and the next correction appears in the $\zeta_1^4$ order
\bea
m^2_U\sim \zeta_1^2+{\mathcal O}(\zeta_1^4)\ .
\eea
Since the coefficient in front of the first term is positive, $m^2_U$ is positive at least for some finite range of $\zeta_1\lesssim 1$. Hence the vacuum states, and the dual gravity backgrounds, are perturbatively stable for $\zeta_1\lesssim 1$.

Besides the perturbative stability, one may also worry about possible tunneling between the $2M$ vacua of the original KS theory with different phases of $\Lambda^3$. When $\zeta_1$ is small, i.e. when  $m_{\lambda\lambda}$ is small, one can estimate the shift of vacuum energy in each of those vacua as \cite{KuSo}:
\begin{eqnarray}
 V\simeq M\Re \left( m_{\lambda\lambda} \Lambda^3 \right)\ .
\end{eqnarray}
Our solutions correspond to $m_{\lambda\lambda}$ aligned with $\Lambda^3$ -- they have the opposite phase such that their product is real. This is so since we have not included in our setup the mode $\lambda(\tau)$ mentioned briefly in footnote \ref{lambda} and its counterpart in the fluxes. It remains unclear, though, whether $m_{\lambda\lambda} \Lambda^3$ is positive or not. Hence our solutions may be unstable non-perturbatively.

Let us show that in any case the theories in question are long-lived, if we do not depart too far from the KS solution.
At the leading order the tension of the domain wall separating various vacua remains the same as in the supersymmetric case \cite{DKS}:
\begin{eqnarray}
T_n = M \left\vert \Lambda \right\vert^3 \left\vert 1-e^{i\pi {n/M}} \right\vert \, , \qquad n=1, \ldots, 2 M \, .
\end{eqnarray}
When $m_{\lambda\lambda}$ is small enough we can use the thin wall approximation and estimate the size of a new vacuum bubble by minimizing the following effective action:
\begin{eqnarray}
S_{\rm bubble}\sim \Delta V R^4 - T_{\Delta n} R^3\ .
\end{eqnarray}
For the adjacent vacua  with $\Delta n=1$ we have $T\sim |\Lambda|^3$ while $\Delta V$ is $1/M$ suppressed.
For a one leap transition into the true vacuum $\Delta n=M$ we get $T\sim M\Lambda^3$ and $\Delta V\sim M m_{\lambda\lambda}\Lambda^3$.
The decay rate for tunneling into a new vacuum with
$\Lambda^3\rightarrow \Lambda^3 e^{i\pi/M}$ is smaller than the decay rate for tunneling into the true vacuum in one leap.
The latter is suppressed by:
\begin{eqnarray}
e^{-S_{\rm bubble}}\sim e^{-M(\Lambda/m_{\lambda\lambda})^3}\ .
\end{eqnarray}
Because of large $M$ for $m_{\lambda\lambda}\lesssim \Lambda$ i.e. $\zeta_1\lesssim 1$ the solutions in question are metastable.

This analysis of stability may be too conservative. In fact it may be that the signs of $m_{\lambda\lambda}$ and $\Lambda^3$ are aligned such that the solutions we find correspond to the true vacuum.

\section{Outline}
\label{sec:outline}

We constructed a one-dimensional family of IIB supergravity solutions which are dual to the $\N=1$ gauge theory perturbed by some combination of relevant single trace and marginal double-trace operators. Thus the family of solutions corresponds to a line in the space of couplings $g_i(\zeta_1)$.
The origin of the line $g_i(0)=0$ corresponds to the original $\N=1$ theory \cite{KS}.
Except for the small region near the origin $\zeta_1\rightarrow 0$ the couplings are  not infinitesimally small and the corresponding operators explicitly break supersymmetry.  Hence our solutions are gravity duals for non-supersymmetric field theories.

The original $\N=1$ theory has only one massless mode which acquires a positive mass, at least for the theories within  a range $\zeta_1  \lesssim 1$. Therefore the solutions in question are perturbatively stable. There is a possible non-perturbative instability associated with tunneling into a vacuum with $\Lambda^3_{\rm new}=-\Lambda^3$. But at least for $\zeta_1\lesssim 1$  the decay rate is exponentially suppressed.

The geometry of the solutions is extremely simple. The unwarped metric on the deformed conifold is Ricci-flat but not K\"ahler for $\zeta_1\neq 0$. The three-form fluxes are ISD and the dilaton is constant. There is also a warp-factor which has a similar behavior to its KS counterpart. The solutions are regular and all fields approach their KS values in UV, up to $1/r$ corrections.

The apparent simplicity of the solutions makes them a natural arena to study strongly coupled dynamics of the non-SUSY confining  gauge theories.
Similarly to the KS background, which was a playground for numerous phenomenological models, our solutions can be used for model-building with an advantage that the resulting models are not supersymmetric.
Because of the GKP structure of the solutions the probe D3-brane in such backgrounds experiences no force. Hence our solutions can be a natural starting place for the models of stringy inflation based on the dynamics of a mobile D3-brane \cite{KKLT,KKLMMT, delicate,explicit, systema, fluxes}. In a similar way the GKP structure makes it possible to use the trick of \cite{DKS7} in its original form and embed the U-shaped D7-branes inside the conifold. Eventually one arrives at a holographic model of baryonic matter with the SUSY broken explicitly by the relevant operators in the gauge sector, and not only by the flavor sector. Moreover because of the KS asymptotic at infinity the resulting baryons will develop a realistic attractive potential at least in some range of parameters \cite{ab}. Finally, it would be interesting to understand the appearance of the flat moduli associated with the mobile D3-branes from the point of view of the dual non-supersymmetric field theory (see \cite{Klebanov:2010tj} for a similar situation in a $3d$ dimensional theory).

\section*{Acknowledgments}

We thank N. Halmagyi for collaboration at the initial stages of this project and G. Giecold and C. Nu$\tilde{\rm n}$ez for reading the manuscript. We are also grateful to I. Klebanov, Z. Komargodski, J. Maldacena, N. Seiberg, M. Shifman and
A. Yung  for numerous discussions.

A.D. thanks the theory group at Institut de Physique Th\'eorique,  CEA-Saclay and High Energy Group at Cornell University for hospitality. The research of A.D. was supported by the DOE grant DE-FG02-90ER40542, by Monell Foundation, and in part by the grant RFBR 07-02-00878 and the Grant for Support of Scientific Schools NSh-3035.2008.2.

The work of S.K. was supported in part by the ERC Starting Independent Researcher Grant 240210 - String-QCD-BH.

S.K. would like to thank Institute for Advanced Study for hospitality and the European Commission Marie Curie Fellowship (under the
contract IEF-2008-237488) for the financial support during the US visit.

\appendix
\section{A. The Ricci flatness equations}

Here we summarize the Ricci-flatness equations for the $6d$ metric (\ref{metric6d})
\begin{eqnarray}
\label{EOMpqyz}
    \ddot{p} + \dot{p} ( 4 \dot{p} + 6 \dot{q} ) &=& - \frac{2}{15} e^{-20 p} \left( 1 + 2 \cosh^2 y \sinh^2 z \right)  \\
&&   \quad   +  \frac{2}{15} e^{-10 p} \cosh y \cosh z + \frac{1}{5} \sinh^2 y \, \nonumber \\
    \ddot{q} + \dot{q} ( 4 \dot{p} + 6 \dot{q} ) &=& - \frac{8}{135} e^{-20 p} \left( 1 + 2 \cosh^2 y \sinh^2 z \right) \nonumber \\
&&      +  \frac{16}{45} e^{-10 p} \cosh y \cosh z - \frac{2}{15} \sinh^2 y \, \nonumber \\
    \ddot{y} + \dot{y} ( 4 \dot{p} + 6 \dot{q} ) &=& \cosh y \sinh y \left(  \frac{8}{9} e^{-20 p} \sinh^2 z -
                \frac{4}{3} e^{-10 p} \frac{\cosh z}{\cosh y} +  \left( 1 + \dot{z}^2 \right) \right) \, , \nonumber
\end{eqnarray}
and:
\begin{equation}
\ddot{z} + \dot{z} ( 4 \dot{p} + 6 \dot{q} + 2 \dot{y} \tanh y )  = \sinh z \left(  \frac{8}{9} e^{-20 p} \cosh z -
                \frac{4}{3} e^{-10 p} \frac{1}{\cosh y}  \right) \, .
\end{equation}
These equations can be derived from the $SU(2) \times SU(2)$ action of \cite{PT} with vanishing fluxes and the relations (\ref{PTtoHERE}):
\begin{eqnarray}
    S & \sim & - \int \d \tau e^{4 p +6 q}  \Bigg[ \left( 15 \dot{p}^2 - \frac{135}{4} \dot{q}^2
                                      + \frac{3}{4} \left( \dot{y}^2 + \cosh^2 (y) \dot{z}^2 \right)  \right)   \\
   && + \left( \frac{1}{3} e^{-20 p} (1 + 2 \cosh^2(y) \sinh^2(z)) - 2 e^{-10 p} \cosh(y) \cosh(z) + \frac{3}{4} \sinh^2 (y) \right) \Bigg]   \, .    \nonumber
\end{eqnarray}
The solutions of these EOM are also subject to the zero energy (zero Hamiltonian) condition following from the action. This condition can be used to find $\dot{q}(\tau)$, and so we are left only with the equations for $p(\tau)$, $y(\tau)$ and $z(\tau)$.

To study the large $\tau$ behavior we have to expand the equations linearly around (\ref{PTtoHERE}).
The equations for $\delta y(\tau)$ and $\delta z(\tau)$ are both homogenous. The $\delta z(\tau)$ equation is solved by $e^{-2\tau/3}$ and $\tau e^{-2\tau/3}$, as expected for the the $\Delta=2$ operator. The two solutions for $\delta y(\tau)$ are $e^{-\tau}$ and $e^{-\tau/3}$ corresponding to $\Delta=3$. Finally, the homogenous part of the equation for $\delta p(\tau)$ gives two modes, $e^{-2 \tau}$ and $e^{2\tau/3}$, implying that the mode has dimension $\Delta=6$.

\bibliographystyle{utphys}
\bibliography{draft}

\end{document}